\documentclass[twocolumn,amssymb,aps,floatfix]{revtex4-2}
\emergencystretch=\maxdimen
\usepackage{graphicx}
\usepackage{dcolumn}
\usepackage{amsmath}
\usepackage[latin1]{inputenc}
\usepackage{amssymb}
\usepackage[colorlinks=true, citecolor=blue, urlcolor = blue, linkcolor= red, bookmarks=true]{hyperref}
\usepackage{float}
\usepackage{amsfonts}
\usepackage{dcolumn}
\usepackage{hyperref}
\usepackage{subfigure}
\usepackage{pgfplots}
\usepackage{booktabs}
\usepackage{mathrsfs}
\usepackage{multirow}
\usepackage{afterpage}
\pgfplotsset{compat=1.16}

\usepackage{wrapfig}
\usepackage{adjustbox}
\makeatletter
\def\btt#1{\texttt{\@backslashchar#1}}
\DeclareRobustCommand\bblash{\btt{\@backslashchar}} \makeatother
\begin{document}
\title{{\bf Investigating strong gravitational lensing effects  by suppermassive black holes with Horndeski gravity}}
\author{Jitendra Kumar $^{a}$ } \email{jitendra0158@gmail.com} 
\author{Shafqat Ul Islam$^{a}$ } \email{Shafphy@gmail.com}

\author{Sushant~G.~Ghosh$^{a,\;b}$} \email{sghosh2@jmi.ac.in, sgghosh@gmail.com}

\affiliation{$^{a}$ Centre for Theoretical Physics, 
	Jamia Millia Islamia, New Delhi 110025, India}
\affiliation{$^{b}$ Astrophysics and Cosmology Research Unit, 
	School of Mathematics, Statistics and Computer Science, 
	University of KwaZulu-Natal, Private Bag 54001, Durban 4000, South Africa}
\begin{abstract}
\begin{center}
{\bf Abstract}
\end{center}
We study gravitational lensing in strong-field limit by a static spherically symmetric black hole in quartic scalar field  Horndeski gravity having additional hair parameter $q$, evading the no-hair theorem. We find an increase in the deflection angle $\alpha_D$, photon sphere radius $x_{ps}$, and angular position $\theta_{\infty}$ that increases more quickly while angular separation $s$ more slowly, but the ratio of the flux of the first image to all other images $r_{mag}$ decreases rapidly with increasing magnitude of the hair $q$.  We also discuss the astrophysical consequences in the supermassive black holes at the centre of several galaxies and note that the black holes in Horndeski gravity can be quantitatively distinguished from the Schwarzschild black hole. Notably, we find that the deviation   $\Delta\theta_{\infty}$  of black holes in Horndeski gravity from their general relativity (GR) counterpart, for supermassive black holes Sgr A* and M87, for $q=-1$ respectively, can reach as much as   $25.192~\mu$as and $18.92~\mu$as  while $\Delta s$ is about  $1.121~\mu$as for Sgr A* and $0.8424~\mu$as for M87*. The ratio of the flux of the first image to all other images suggest that the Schwarzschild images are brighter than those of the black holes in Horndeski gravity, wherein the deviation  $|\Delta r_{mag}|$ is as much as 3.082.  The results suggest that observational tests of hairy black holes in Horndeski gravity are indeed feasible. 
\end{abstract}

\pacs{ } \maketitle
\section{Introduction}
Gravitational lensing by black holes is one of the most powerful astrophysical tools for investigating the strong-field features of gravity and provide us with information about the distant stars that are too dim to be observed. It can help us detect exotic objects and hence verify alternative theories of gravity.  The gravitational lensing theories were developed, among others, by  Liebes \cite{Liebes:1964}, Refsdal \cite{Refsdal:1964}, and Bourassa and Kantowski \cite{Bourassa:1973}.   They have successfully explained the astronomical observations but in the weak field approximation.  However, when a lens is a compact object with a photon sphere (such as a black hole), a strong field treatment of gravitational lensing is needed instead because photons passing close to the photon sphere have large deflection angles.  Virbhadra and Ellis \cite{Virbhadra:1999nm} obtained the lens equation using an asymptotically flat background metric,  in the strong-field limit for a  Schwarzchild black hole numerically. Apart from the primary and secondary images, they reported two infinite sets of faint relativistic images. An exact lens equation without reference to a background metric was found by Fritelli et al. \cite{Frittelli:1998hr}. Later, Bozza \cite{Bozza:2002zj} used the strong field limit approximation to obtain analytical expressions for the positions and magnification of the relativistic images and extended his  method of lensing for a general class of static and spherically symmetric spacetimes to show that the logarithmic divergence of the deflection angle at photon sphere is a generic feature for such spacetimes.   Bozza's  \cite{Bozza:2002zj}  methods was extended to several static, spherically symmetric metrics which includes  Reissner$-$Nordstrom black holes \cite{Eiroa:2003jf}, braneworld black holes \cite{Eiroa:2004gh,Whisker:2004gq,Eiroa:2005vd,Li:2015vqa}, charged black hole of heterotic string theory \cite{Bhadra:2003zs}.  The strong gravitational field continues to receive significant attention,  more recent works include lensing from other black holes \cite{Chen:2009eu,Sarkar:2006ry,Javed:2019qyg,Shaikh:2019itn} and from various modifications of Schwarzschild geometry \cite{Eiroa:2010wm,Ovgun:2019wej,Panpanich:2019mll,Bronnikov:2018nub,Shaikh:2018oul, Lu:2021htd,Babar:2021nst}, and more recently in $4D$ Einstein-Gauss-Bonnet gravity \cite{Kumar:2020sag,Islam:2020xmy,Narzilloev:2021jtg}.  The gravitational lensing by a primary photon sphere with unstable circular light orbits and by a secondary photon sphere on a wormhole throat in a black-bounce regular spacetime shows the existence of an antiphoton sphere and the formation of infinite images near it \cite{Tsukamoto:2021caq}. The gravitational lensing received a boost when Event Horizon Telescope (EHT) \cite{Akiyama:2019cqa, Akiyama:2019bqs} unravelled the first-ever image of the supermassive black hole M87*. These results offer testing grounds for gravity theories on offering a compelling probe of the strong gravitational fields.  With this motivation, this paper investigates the strong-field gravitational lensing of light by the hairy black holes in Horndeski gravity \cite{Bergliaffa:2021diw}. 

 As a modification to GR, the simplest extensions are the scalar-tensor theories like Horndeski gravity \cite{Horndeski:1974wa}, probably the most general four-dimensional scalar-tensor theory with equations of motion containing up to second-order derivatives of the dynamical fields.  Horndeski theory of gravitation is described by the action principle formulated from the metric and a scalar field that leads to field equations  with no derivatives beyond second order for the metric and the scalar field, and the theory has the same symmetries as GR, namely, diffeomorphism and local Lorentz invariance.   (see e.g. \cite{Damour:1992we,Horbatsch:2015bua}).  All the terms present in the action of Horndeski gravity have been shown to be originating from Galileons, i.e. scalar-tensor models having Galilean symmetry in flat spacetime~\cite{Nicolis:2008in}.  There are compelling arguments that suggest that certain modifications are required in GR at both very high and very low energy scales. Gravitational collapses are destined to unavoidable singularities, while on cosmological scales, to describe the observed accelerated expansion of the Universe, GR relies on the yet unexplained presence of dark energy \cite{Clifton:2011jh}. Horndeski theories \cite{Kobayashi:2019hrl} which have been studied in both the strong gravity on compact objects, such as neutron stars, black holes \cite{Maselli:2016gxk} and in cosmological regimes to describe the accelerated expansion \cite{Kase:2018aps}. The space of solutions for Horndeski's theory of gravity is endowed with hairy black holes \cite{Rinaldi:2012vy,Babichev:2014fka,Babichev:2017guv,Anabalon:2013oea,Cisterna:2014nua,Bravo-Gaete:2014haa},  Among the static and spherically symmetric hairy black holes in scalar-tensor theories  the  simplest case in which solutions admits a hairy profile with a radially dependent scalar field was studied in \cite{Sotiriou:2013qea,Sotiriou:2014pfa,Babichev:2016rlq,Benkel:2016rlz,Babichev:2017guv}. The   time-dependent hairy black hole solutions within the Horndeski class of theories have also been obtained \cite{khoury}. In  \cite{Hui:2012qt}, authors provide a no-hair theorem for static and sphericaly symmetric black hole solutions with vanishing  Galileon hair at infinity which has further examined by Babichev {\it et. al.} \cite{Babichev:2017guv} considering Horndeski theories and beyond it. They demonstrated that shift-symmetric Horndeski theories including the extended ones  allow for static and asymptotically flat black holes with a static scalar field \cite{Babichev:2017guv}  such that the Noether current associated with shift symmetry vanishes, while the scalar field cannot be trivial; In turn, it leads to hairy black holes for the quartic Horndeski gravity \cite{Babichev:2017guv}.   Lately, the investigation of black holes in Horndeski and beyond Horndeski theories has received significant attention \cite{hbh,Bergliaffa:2021diw}.

We investigate the predictions of spherical hairy black holes in quartic Horndeski gravity  \cite{Bergliaffa:2021diw} for the strong-field gravitational lensing effects of supermassive black holes at the center of the Milky Way and other galaxies. Our most exciting result is that the difference between the angular positions of relativistic primary and secondary images in Horndeski gravity and GR could be as large as $\mu$as. Also, the calculated  values of time delay between these images are different in GR and Horndeski gravity, and the difference could be as significant as seconds. These suggest that observational tests of Horndeski gravity are indeed feasible.

The paper is organized as follows: We begin with briefly reviewing the hairy black holes in Horndeski gravity in the Sec. \ref{sec2}. Restrictions on parameters from the horizon structure and deflection of light is the subject of Sec. \ref{sec3}. Moreover, we also discuss the strong lensing observables by the hairy black holes, including the image positions $\theta_{\infty}$, separation $s$, magnifications $\mu_n$  in Sec. \ref{sec3}.  Time delay between the first and second image when they are on the same side of source have been calculated for supermassive black holes SgrA*, M87* and  those at the centers of 21 other galaxies in Sec. \ref{sec4}.  A numerically analysis of  the observables by  taking the supermassive black holes NGC 4649,  NGC 1332, Sgr A* and M87*, as the lens is part of  Sec. \ref{sec5}. Finally, we summarize our results to end the paper in Sec. \ref{sec6}. 

We will work in units where $G = c = 1$.

\section{Hairy Black Holes in Horndeski Theory of Gravity}\label{sec2}
Horndeski gravity is described by the action formulated from the metric and the scalar field \cite{Babichev:2017guv}. It involves $4$ arbitrary functions $Q_i$ ($i=2,..5$) of kinetic term $\chi =-{\partial^\mu \phi \partial_\mu \phi}/{2}$ \cite{Babichev:2017guv}. Here, we have considered the particular type of the action of \cite{Babichev:2017guv} which is quartic, i.e, $Q_5$ term is absent $(Q_5=0)$. The hairy black hole solution we are interested is derived from the quartic Horndeski gravity \cite{Bergliaffa:2021diw} whose action reads
\begin{align}
	\label{eq0}
	S=&\int d^4x \sqrt{-g}\Big\{Q_2(\chi)+Q_3(\chi)\square\phi+ Q_4(\chi)R\\
	+& Q_4,_\chi[(\square\phi)^2-(\nabla^\mu\nabla^\nu\phi)(\nabla_\mu\nabla_\nu\phi)]\Big\},
\end{align}
where $g \equiv \text{det}(g_{\mu\nu})$, $g_{\mu\nu}$ is the metric tensor, $R$ and $G_{\mu\nu}$, respectively, denote Ricci scalar and Einstein tensor. The  $\square$ is the d'Alembert operator and $\nabla_\mu$ is the covariant derivative.  The $4$-current vector associated with the Noether charge is \cite{Bergliaffa:2021diw}, 
$$j^\nu=\frac{1}{\sqrt{-g}}\frac{\delta S}{\delta(\phi_{,\mu})},$$ which results into
\begin{eqnarray}
	\label{eQ_2}
	\nonumber
	j^\nu=-Q_2,_\chi \phi^{, \nu}-Q_3,_\chi (\phi^{, \nu}\square\phi+\chi^{, \nu})~~~~~~~~~~~~~~~~~~~~\\
	-Q_4,_\chi (\phi^{, \nu}R-2R^{\nu\sigma}\phi,_\sigma)~~~~~~~~~~~~~~~~~~~~~~~~~~~\\
	\nonumber
	-Q_4,_\chi,_\chi\{\phi^{, \nu}[(\square \phi)^2
	-(\nabla_\alpha\nabla_\beta\phi)(\nabla^\alpha\nabla^\beta\phi)]\\
	\nonumber
	+2(\chi^{, \nu}\square\phi-\chi,_\mu\nabla^{\mu}\nabla^{\nu}\phi)
	\},
\end{eqnarray} 
where we have used 
the usual convention for the Riemann tensor
\begin{eqnarray}
	\label{eq1}
	\nabla_\rho\nabla_\beta\nabla_\alpha\phi-\nabla_\beta\nabla_\rho\nabla_\alpha\phi=-R^\sigma_{~\alpha\rho\beta}\nabla_\sigma\phi.
\end{eqnarray}
Varying the action (\ref{eq0}) with respect to metric tensor $g^{\mu\nu}$ we obtain the field equations \cite{Bergliaffa:2021diw, Babichev:2017guv}
\begin{eqnarray}
	\label{eq3}
	Q_4 G_{\mu\nu}= T_{\mu\nu},
\end{eqnarray}
where
\begin{eqnarray}
	\label{eq4}
	\nonumber
	T_{\mu\nu}=\frac{1}{2}(Q_2,_\chi \phi ,_\mu \phi ,_\nu+Q_2 g_{\mu\nu})
	+\frac{1}{2}Q_3,_\chi(\phi ,_\mu \phi ,_\nu\square\phi~~~\\
	\nonumber
	-g_{\mu\nu} \chi,_\alpha \phi^{, \alpha}+\chi,_\mu \phi ,_\nu
	+\chi,_\nu \phi ,_\mu)- Q_4,_\chi\Big\{\frac{1}{2}g_{\mu\nu}[(\square\phi)^2\\
	\nonumber
	-(\nabla_\alpha\nabla_\beta\phi)(\nabla^\alpha\nabla^\beta\phi)-2R_{\sigma\gamma}\phi^{,\sigma}\phi^{,\gamma}]
	-\nabla_\mu\nabla_\nu \phi \square\phi\\
	\nonumber
	+\nabla_\gamma\nabla_\mu \phi \nabla^\gamma \nabla_\nu \phi-\frac{1}{2}\phi ,_\mu \phi ,_\nu R 
	+R_{\sigma\mu}\phi^{,\sigma}\phi,_{\nu}\\
	+R_{\sigma\nu}\phi^{,\sigma}\phi,_{\mu}+R_{\sigma\nu\gamma\mu} \phi^{,\sigma}\phi^{,\gamma}
	\Big\}~~~~~~~~~~~~~~\\
	\nonumber
	-Q_4,_\chi,_\chi \Big\{g_{\mu\nu}(\chi,_{\alpha}\phi^{,\alpha}\square \phi+\chi_{,\alpha} \chi^{, \alpha})+\frac{1}{2}\phi ,_\mu \phi ,_\nu\times
	\\
	\nonumber
	(\nabla_\alpha\nabla_\beta\phi\nabla^\alpha\nabla^\beta\phi-(\square\phi)^2)
	- \chi,_\mu \chi,_\nu \\
	\nonumber
	- \square\phi( \chi,_\mu \phi,_\nu  
	+ \chi,_\nu \phi,_\mu)
	\\
	\nonumber
	- \chi,_\gamma[\phi^{,\gamma}\nabla_\mu\nabla_\nu\phi-(\nabla^\gamma\nabla_\mu\phi)\phi,_{\nu}
	-(\nabla^\gamma\nabla_\nu\phi)\phi,_{\mu}]
	\Big\}.
\end{eqnarray} 
\begin{figure*} 
	\begin{centering}
		\begin{tabular}{cc}
		    \includegraphics[scale=0.82]{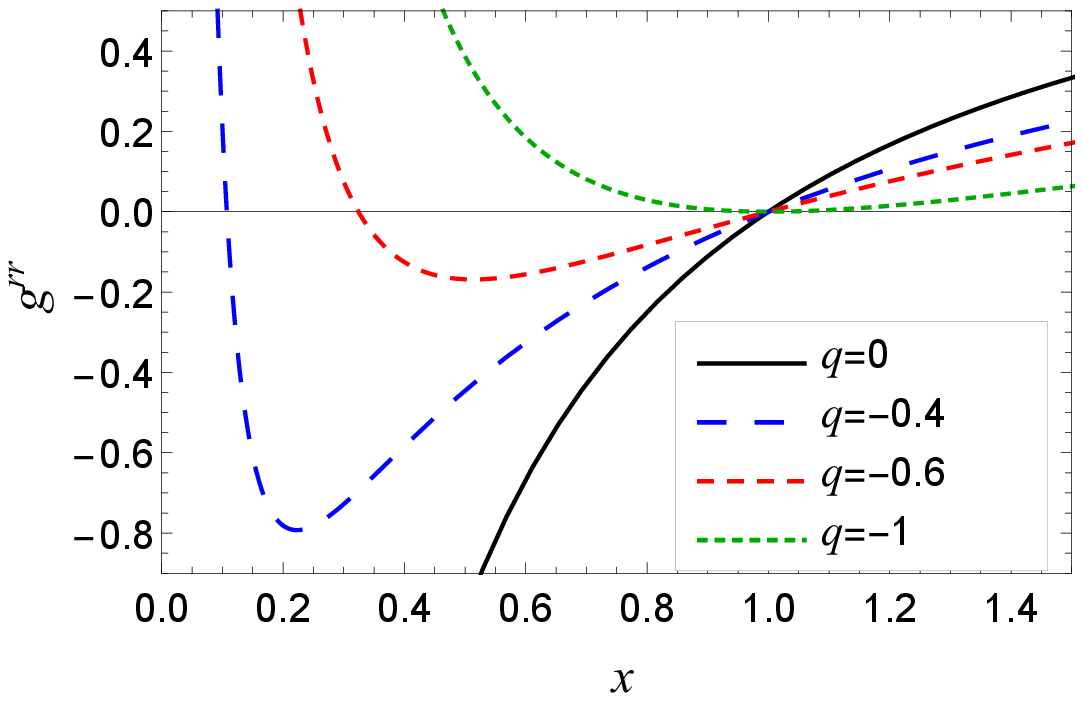}
		    \includegraphics[scale=0.82]{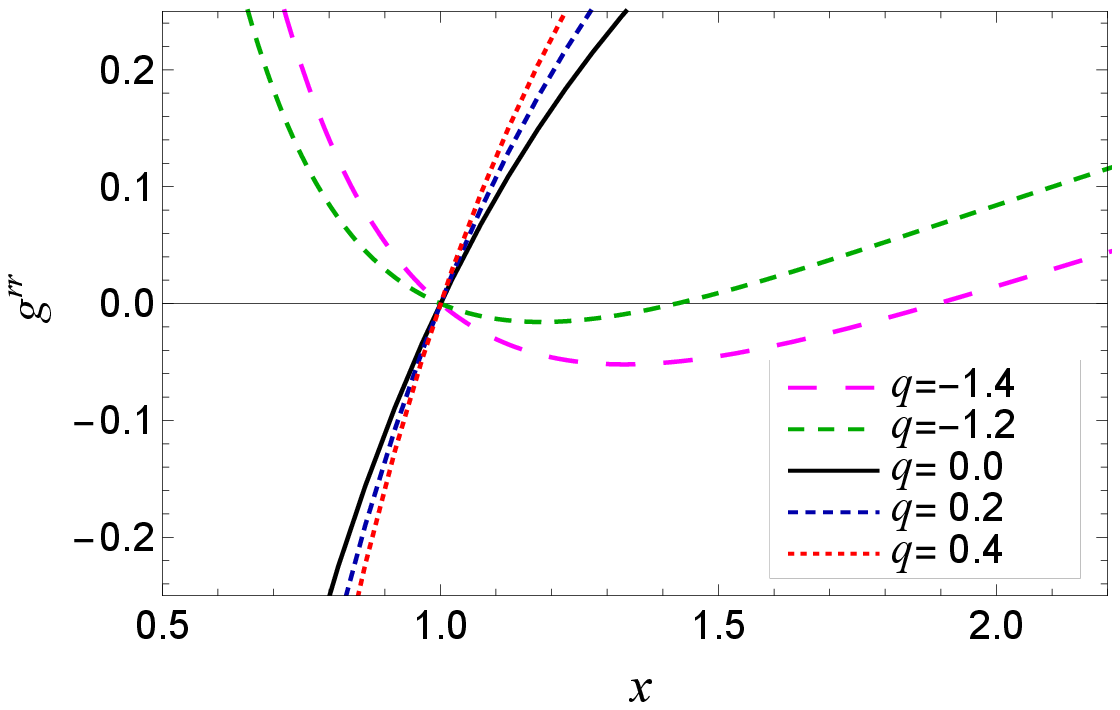}\\
		   \end{tabular}
	\end{centering}
	\caption{The horizons of the hairy black hole in Horndeski gravity. When $-\infty < q < 0$ the black hole admits both event horizon $x_{+}$ and Cauchy horizon $x_{-}$, otherwise has only  event horizon for $q>0$. }\label{fig1}
\end{figure*}
Henceforth, we shall specialise to a scalar field $\phi\equiv\phi(r)$ \cite{Bergliaffa:2021diw, Babichev:2017guv}. It will be the source of a static and spherically symmetric spacetime. The integration of field equations leads to the black hole solution \cite{Bergliaffa:2021diw}
\begin{eqnarray}\label{eq16}
	ds^2=-A(r)dt^2+\frac{1}{B(r)}dr^2+r^2(d\theta^2+\sin^2\theta d\varphi^2),
\end{eqnarray}
where
\begin{eqnarray}
	\label{eqf}
	A(r)=B(r)=1 -\frac{2m}{r} + \frac{q}{r}\ln\left({\frac{r}{2m}}\right),
\end{eqnarray}
where $m$ is the integration constant related to the black hole mass and  $q$ is a constant that results from Horndeski gravity. The metric (\ref{eq16}) represents a hairy black hole and encompasses the Schwarzschild metric in the limit $q \to 0$. The solution (\ref{eq16}) is asymptotically flat since
$\displaystyle{\lim_{r \to \infty}}$ $ A(r)=B(r)=1$. The Kretschmann and Ricci scalars diverge \cite{Bergliaffa:2021diw} along $r=0$, establishing the metric (\ref{eq16}) is scalar polynomial singular.

\section{Strong Gravitational Lensing by hairy black hole in Horndeski gravity}\label{sec3}
In this section we shall study gravitational lensing by hairy black hole (\ref{eq16}) to investigate how the parameter $q$ affects the lensing observables in strong field limit. For our study of lensing we shall restrict the value of $q$ in the range $-1 \leq q \leq 0$. It should give us useful insights about the possible effects of the parameter $q$ on strong gravitational lensing. It is convenient to measure quantities $r,q,t$ in terms of the Schwarzschild radius $2m$ \cite{Bozza:2002zj} and use $x$ instead of $r$, to rewrite the metric~(\ref{eq16}) as
\begin{equation}\label{metric}
ds^2=-A(x)dt^2+\frac{1}{B(x)}dx^2+C(x)\Big(d\theta^2+\sin^2\theta
d\phi^2\Big),
\end{equation}
where
\begin{equation}\label{metrcomp}
A(x)=B(x)=1- \frac{1}{x}+\frac{q}{x}\ln(x),\;\;\;\; C(x)=x^2.
\end{equation}
In addition to curvature scalar polynomial singularity at $x=0$ the metric (\ref{metric}) is also singular at points where $B(x)=0$, which are coordinate singularities and the corresponding surfaces are called horizons. 
In the domain $-1 \leq q < 0$, a simple root analysis of $B(x)=0$ implies existence of two positive roots ($x_{\pm}$), corresponding to the Cauchy ($x_{-}$) and event horizon ($x_{+}$), given by 
\begin{equation}
x_{-}=q~ \text{ProductLog}~\left[\frac{\text{exp}~(1/q)}{q}\right], \;\;
x_{+}=1,
\end{equation}
 where $\text{ProductLog}~(z)$, for arbitrary $z$, is defined as the principal solution of the equation $w~\text{exp}~(w)=z$.

\begin{figure} 
		   \includegraphics[scale=0.8]{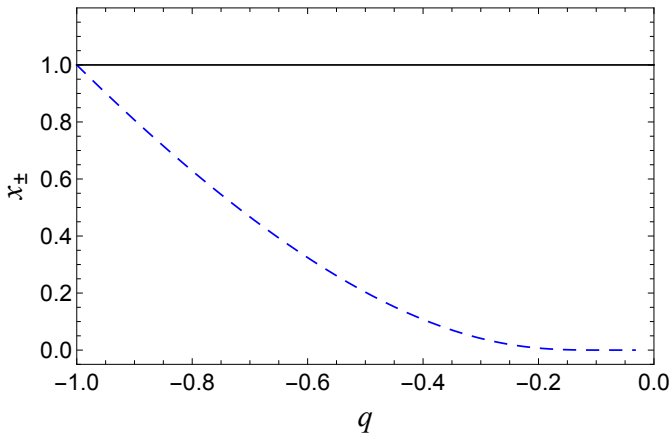}
	\caption{ The  event horizon $x_{+}$ (solid line) and Cauchy horizon $x_{-}$ (dashed line) for black holes in Horndeski gravity.}\label{fig2}
\end{figure}

The metric (\ref{metric}) always has a horizon at $x=1$, irrespective of the value of the parameter $q$.  For $0 \leq q < \infty$, the metric (\ref{metric}) has only one horizon namely event horizon fixed at  radius $x=1$ (cf. Fig.~\ref{fig1}). For $-1 \leq q < 0$, unlike the Schwarzschild spacetime, it displays two horizons (Cauchy and event)  (cf. Fig.~\ref{fig1}). The event horizon ($x_{+}$) is fixed at the  radius $x=1$, while Cauchy horizon ($x_{-}$) increases with decreasing $q$ and merges with the event horizon ($x_{+}$) in the limit $q \to -1$ (cf. Fig.~\ref{fig2}). Since the metric (\ref{metric}) has curvature singularity at $x=0$, the existence of horizon for any value of $q$ means the cosmic censorship hypothesis \cite{Penrose:1969pc} is respected for hairy black hole in Horndeski gravity.

The strong field gravitational lensing is governed by deflection angle and lens equation. For this, we first observe that a light-like geodesic of the metric (\ref{metric}) admits two constants of motion, namely the energy $\mathcal{E}=-p_{\mu}\xi^{\mu}_{(t)}$ and angular momentum $\mathcal{L}=p_{\mu}\xi^{\mu}_{(\phi)}$, where $\xi^{\mu}_{(t)}$ and $\xi^{\mu}_{(\phi)}$ are, respectively, the Killing vectors due to time-translational and rotational invariance. The null geodesic equation satisfies $ds^2=0$, which gives 
\begin{equation}\label{Veff}
\left(\frac{dx}{d\tau}\right)^2\equiv \dot{x}^2={\cal E}^2-\frac{\mathcal{L}^2A(x)}{C(x)}.
\end{equation}
The static and spherically symmetric compact objects with a strong gravitational field in general relativity have circular photon orbits called photon spheres. The photon sphere being one of the crucial character for strong gravitational lensing \cite{Bozza:2002zj,Virbhadra:1999nm} does not evolve with time, or in other words, null geodesic initially tangent to the photon sphere hypersurface remains tangent to it. The radius of photon sphere, $x_{ps}$ is the greatest positive solution  of the equation \cite{Claudel:2000yi,Virbhadra:2002ju}
\begin{equation}
\frac{C'(x)}{C(x)}=\frac{A'(x)}{A(x)}\label{root} 
\implies x_{ps}=\frac{3q}{2}~\text{ProductLog}\left[\frac{2\text{exp}~(\frac{1}{3}+\frac{1}{q})}{3q}\right].
\end{equation}
From Fig.~\ref{fig4}, we observe that when $q \to 0$, we recover the photon sphere radius, $x_{ps}=1.5$ for the Schwarzschild black hole spacetime \cite{Bozza:2002zj}. By solving $V_{\text{eff}}(x_0)=0$ and recognising the ratio $\mathcal{L}/\mathcal{E}$ as the impact parameter, we get the expression for impact parameter $u$ in terms of the closest approach distance $x_0$ as follows \cite{Bozza:2002zj}
\begin{equation}\label{impact}
u\equiv \frac{\cal L}{\cal E} =\sqrt{\frac{C(x_0)}{A(x_0)}}.
\end{equation} 
The radial effective potential  from Eq. (\ref{Veff}), takes the form
\begin{eqnarray}\label{veff}
\frac{V_{\text{eff}}(x)}{{\cal E}^2} &=&  \frac{u^2}{x^2}\left[1- \frac{1}{x}+\frac{q}{x}\ln(x)\right]-1,
\end{eqnarray}
which describes different kinds of possible trajectories. Photons, coming from the far distance source, approach the black hole with some impact parameter and get deflected symmetrically to infinity, meanwhile reaching a minimum distance ($x_0$) near the black hole. 
\begin{figure*} 
	\begin{centering}
		\begin{tabular}{cc}
		    \includegraphics[scale=0.8]{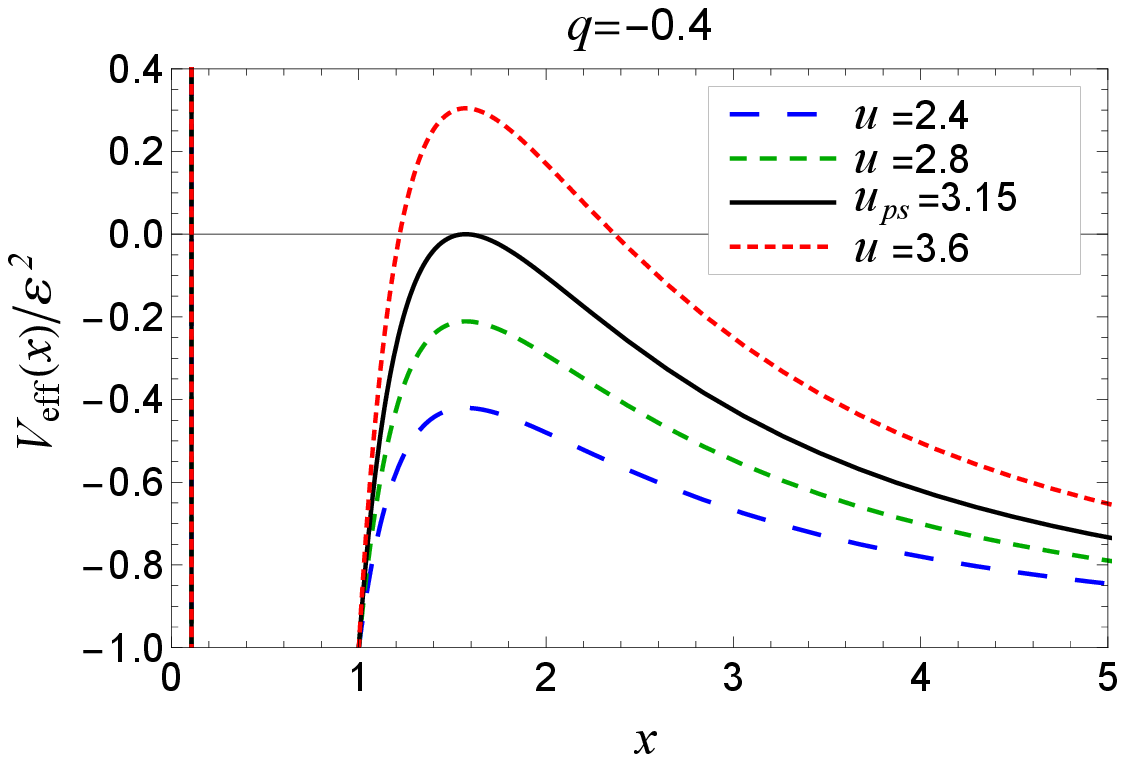}
		    \includegraphics[scale=0.8]{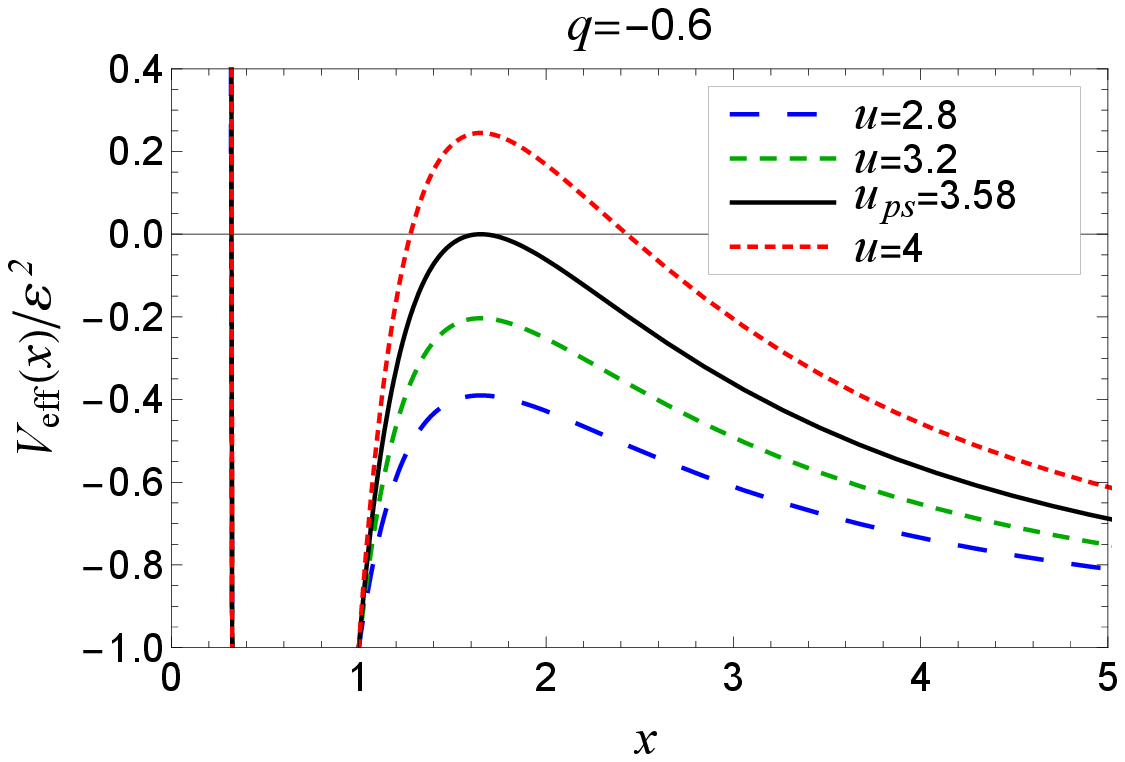}
			\end{tabular}
	\end{centering}
	\caption{Variation of the effective potential $V_{\text{eff}}$ as a function of  radial coordinate $x$, for different values of $q$ and $u$. The photons with critical impact parameter ($u_{ps}$) (black solid curve) make unstable circular orbits.}\label{fig3}	
\end{figure*}
 It turns out that light ray exist in the region where 
$V_{\text{eff}}(x) \leqslant 0$ (cf. Fig.~\ref{fig3}). Further, one can define an unstable (or a stable circular orbit) satisfying  $V_{\text{eff}}(x)=V'_{\text{eff}}(x)=0$ and  
$V''_{\text{eff}}(x_{ps}) < 0$ (or $V''_{\text{eff}}(x_{ps}) > 0$). Next, the first and second derivative of $V_{\text{eff}}(x)$ are 

\begin{equation}\label{veffprime}
\frac{V'_{\text{eff}}(x)}{{\cal E}^2} =  \frac{u^2}{x^4}\left[3+q-2x-3q\ln(x)\right],
\end{equation}

\begin{equation}\label{veffdprime}
\frac{V''_{\text{eff}}(x)}{{\cal E}^2} =  -\frac{u^2}{x^5}\left[12+7q-6x-12q\ln(x)\right].
\end{equation}
For the hairy black hole in Horndeski gravity we find that $V''_{\text{eff}}(x_{ps}) < 0$, which corresponds to the unstable photon circular orbits (cf. Fig.~\ref{fig3}). These photon circular orbits are unstable \cite{Chandra:1992pc} against small radial perturbations, which would finally drive photons into the black hole or toward spatial infinity. 

The deflection angle becomes unboundedly large at $x_0=x_{ps}$ and is finite only for $x_0>x_{ps}$.
The critical impact parameter $u_{ps}$ is defined as
\begin{equation}\label{crimp}
u_{ps}=\sqrt{\frac{C(x_{ps})}{A(x_{ps})}},
\end{equation} 
and depicted in Fig.~\ref{fig4}. The photons with impact parameter $u < u_{ps}$ fall into the black hole, while photons with impact parameter $u > u_{ps}$, reaching the minimum distance $x_0$ near the black hole, are scattered to infinity. The photons only with impact parameter exactly equal to the critical impact parameter $u_{ps}$ revolve around the black hole in unstable circular orbits and generate a photon sphere of radius $x_{ps}$. 

The deflection angle for the spacetime (\ref{metric}) is given by \cite{Claudel:2000yi,Virbhadra:2002ju}
\begin{align}\label{alpha}
\alpha_D(x_0)=I(x_0)-\pi= 2\int^{\infty}_{x_0}\frac{\sqrt{B(x)}dx}{\sqrt{C(x)}
\sqrt{\frac{C(x)A(x_0)}{C(x_0)A(x)}-1}}-\pi,
\end{align}
where $x_0$ is the closest approach distance of the winding photon. Following Bozza \cite{Bozza:2002zj,Chen:2009eu}, we define a variable $z=1-x_0/x$ and exploring the relation between the impact parameter $u$ and closest approach distance $x_0$ in Eq. (\ref{impact}), we find the deflection angle in strong field limit yields
\begin{equation}\label{def}
\alpha_D(u)=-\bar{a}\log{\bigg(\frac{u}{u_{ps}}-1\bigg)}+\bar{b} + \mathcal{O}(u-u_{ps}) \log(u-u_{ps}),
\end{equation}
where $u \approx \theta D_{OL}$. The coefficients  $\bar{a}$ and $\bar{b}$ for the case of hairy black hole in Horndeski gravity are given by
\begin{eqnarray}
\bar{a} = \frac{1}{\sqrt{1+\frac{3}{2}\frac{q}{x_{ps}}}},\;
\bar{b} =-\pi+b_R+\bar{a}\log\left[{\frac{2p_2(x_{ps})}{A(x_{ps})}}\right],\\
b_R = \int_{0}^{1} [R(z,x_{ps})f(z,x_{ps})-R(0,x_{ps})f_0(z,x_{ps})]dz,\nonumber\\
\end{eqnarray}
\begin{eqnarray}
&R(z,x_0)=\frac{2x^2\sqrt{ABC_0}}{x_0C}=2,~~
f(z,x_0) = \frac{1}{\sqrt{A_0-A\frac{C_0}{C}}},\;\;\\
&f_0(z,x_0) = \frac{1}{\sqrt{p_1(x_0)z+p_2(x_0)z^2}},\\
&p_1(x_0) = \frac{3q\log{(x_0)}+2x_0-q-3}{x_0},\\
&p_2(x_0)=\frac{-6q\log{(x_0)}-2x_0+5q+6}{2x_0}.
\end{eqnarray}

\begin{figure*}[t] 
	\begin{centering}
		\begin{tabular}{c c}
		    \includegraphics[scale=0.85]{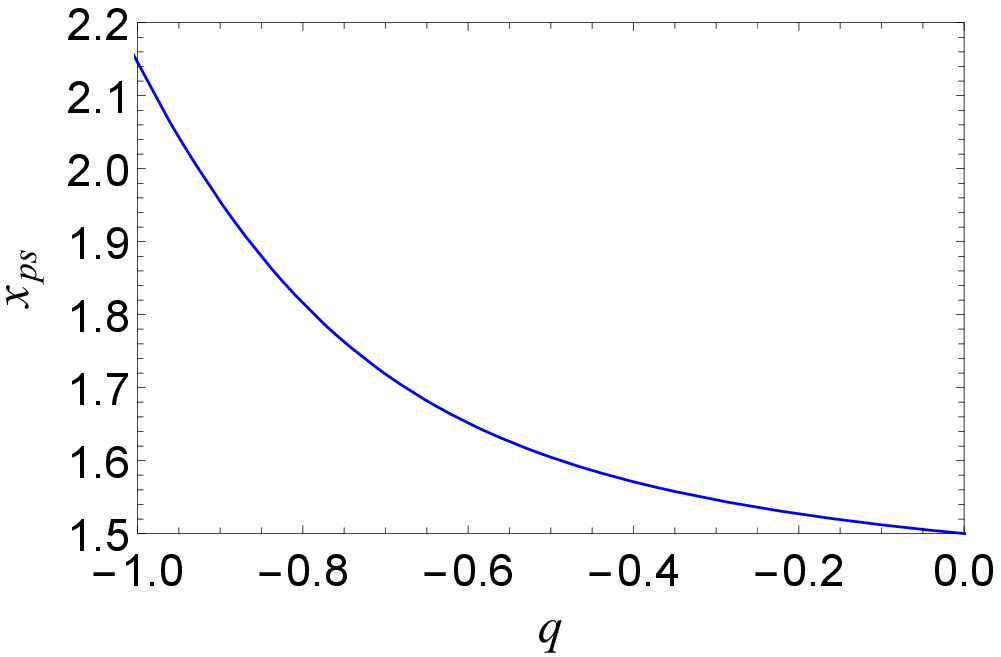}&
		    \includegraphics[scale=0.85]{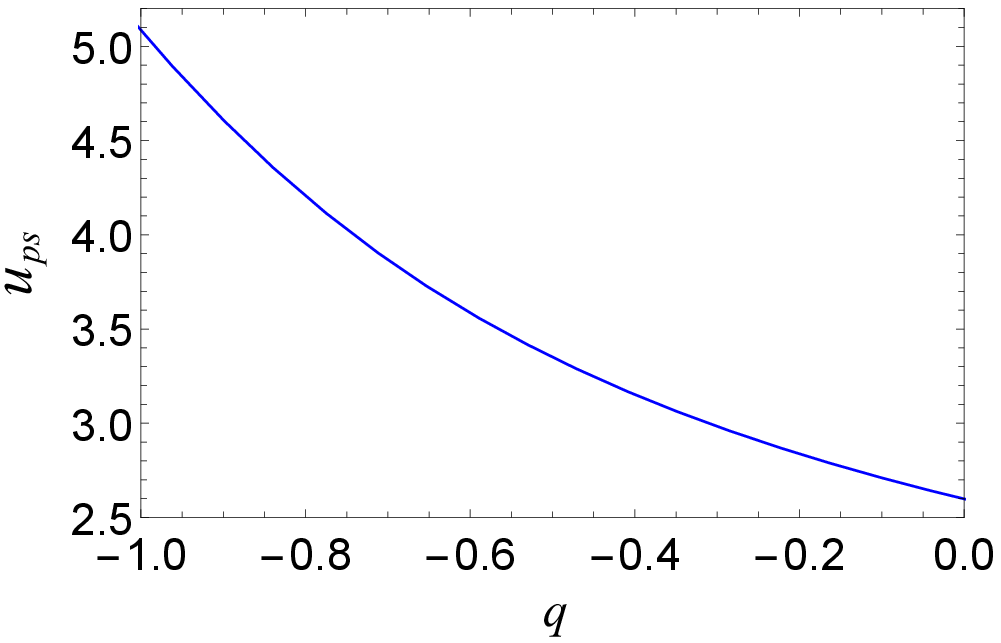}
		    \end{tabular}
	\end{centering}
	\caption{The behavior of the photon sphere radius $x_{ps}$ (Left) and the critical impact parameter $u_{ps}$ (Right) as a function of the hair parameter $q$. As $q \to 0$, the values  $x_{ps} \to 1.5$ and $u_{ps} \to 2.598$ correspond to the Schwarzschild black hole. }\label{fig4}
\end{figure*} 

\begin{figure*}[t] 
	\begin{centering}
		\begin{tabular}{cc}
		    \includegraphics[scale=0.7]{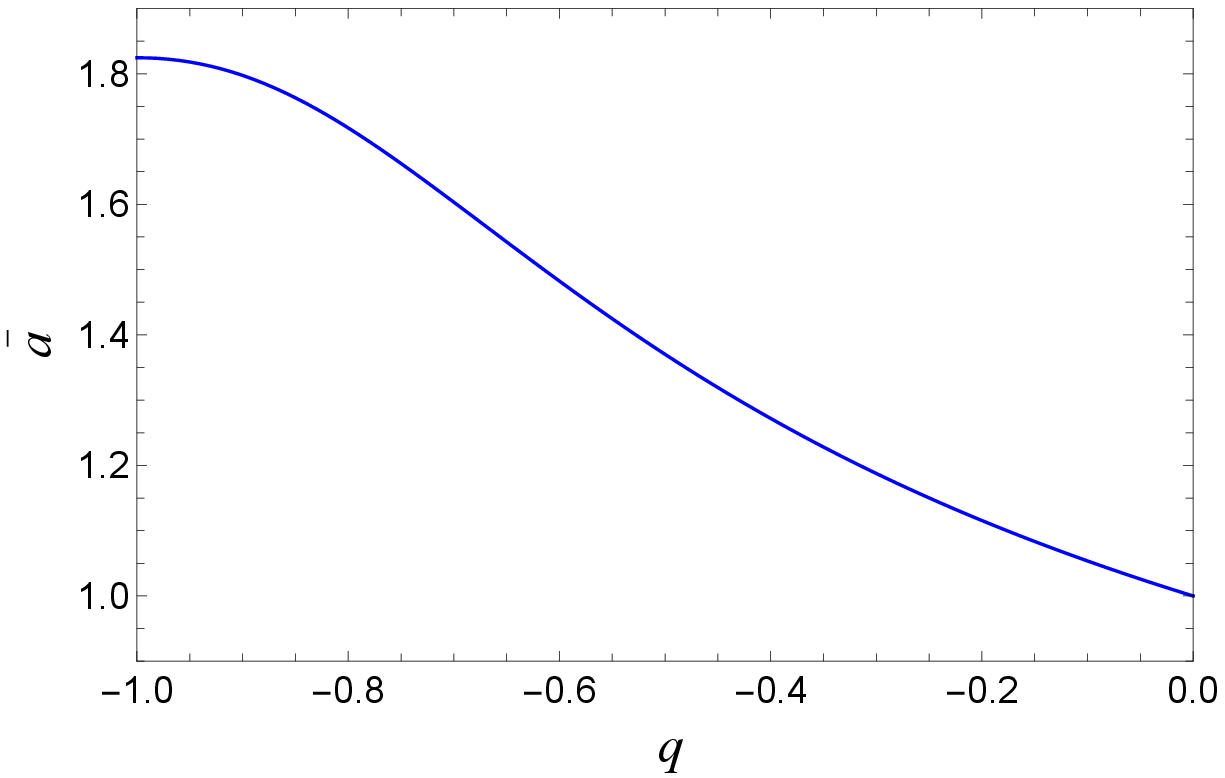}
		    \includegraphics[scale=0.7]{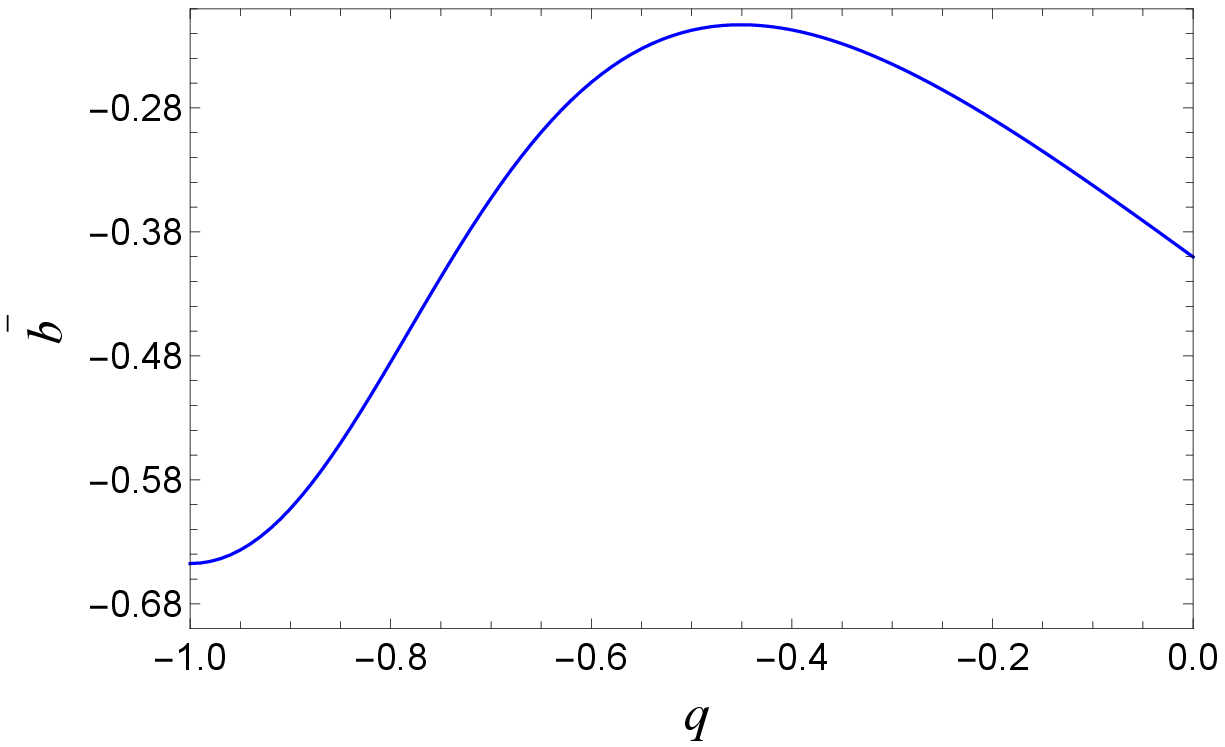}
			\end{tabular}
	\end{centering}
	\caption{The behavior of strong lensing coefficients $\bar{a}$ and $\bar{b}$ as a function of the hair parameter $q$. $\bar{a}=1$ and $\bar{b}=-0.4002$ at $q=0$ correspond to the values Schwarzschild black hole.}\label{fig5}	
\end{figure*}

\begin{figure*}[t] 
	\begin{centering}
		\begin{tabular}{cc}
		    \includegraphics[scale=0.77]{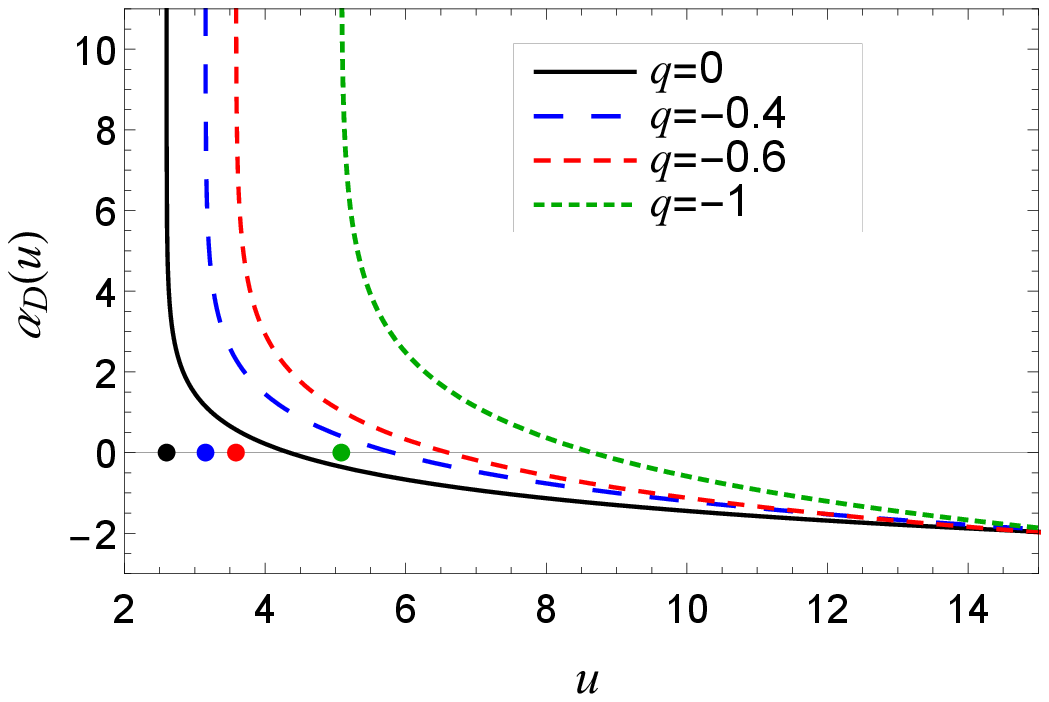}
		    \includegraphics[scale=0.86]{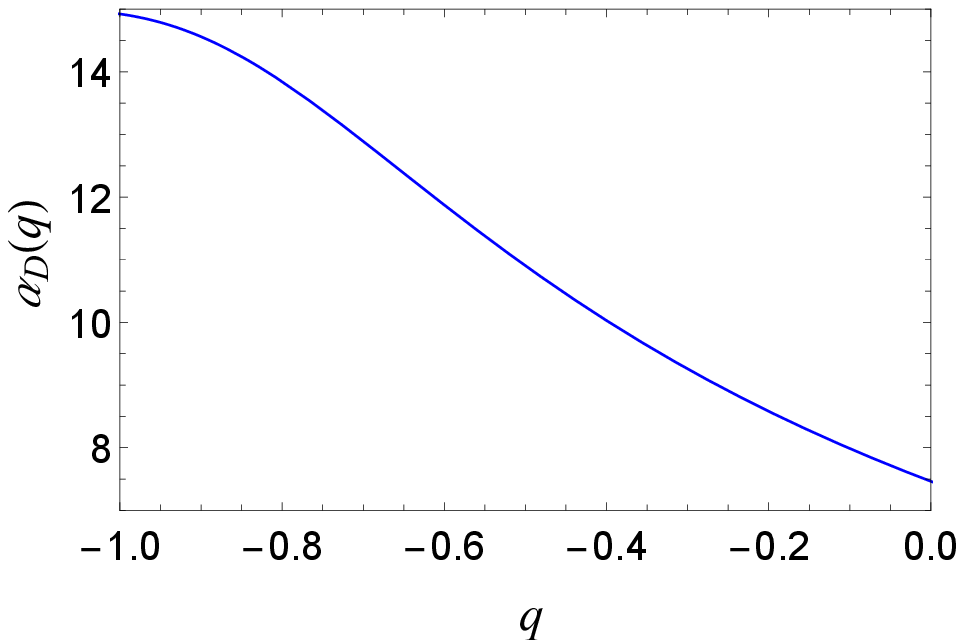}
			\end{tabular}
	\end{centering}
	\caption{(a) (Left) The Variation of deflection angle as a function of  impact parameter $u$ for different values of the parameter $q$. Points on the horizontal axis represent the values of the impact parameter $u=u_{ps}$ at which the deflection angle diverges. (b) (Right) Deflection angles evaluated at $u=u_{ps}+0.001$ as function of the parameter $q$.} \label{fig6}		
\end{figure*}

The deflection angle for the hairy black hole (\ref{metric}) is depicted in Fig. \ref{fig6}, which is monotonically decreasing and $\alpha_D \to \infty$ as $u \to u_{ps}$. When compared with the Schwarzschild black hole  (cf. Fig. \ref{fig6}), the deflection angle for the hairy black hole increases with the increasing magnitude of $q$. A light ray whose $u$ is close enough to $u_{ps}$ can pass close to the photon sphere and go around the lens once, twice, thrice, or many times before reaching the observer. Thus strong gravitational field,  in addition to the primary and secondary images, can give a large number (theoretically an infinite sequence) of images on both sides of the optic axis, which is the line joining the observer and the lens. The two infinite sets of relativistic images correspond to clockwise winding around the black hole and the other produced by counterclockwise winding.  When $x_0 \approx x_{ps}$, the coefficient $p_1(x_0)$ vanishes and the leading term of the divergence in $f_0(z,x_0)$ is $z^{-1}$~\cite{Bozza:2002zj}, thus the integral  diverges logarithmically. The coefficient $b_R$ is evaluated numerically. The coefficient $\bar{a}$ decreases  while $\bar{b}$ increases at first and then, reaching its maximum at $q \approx -0.45$ (cf. Fig. \ref{fig5}), decreases.  The coefficients $\bar{a}=1$ and $\bar{b}=-0.4002$ \cite{Bozza:2002zj}  correspond to the case of the Schwarzschild black hole (cf. Table~\ref{table1}).

\begin{table}
\centering
\begin{tabular}{p{2cm} p{2cm} p{2.5cm} p{1cm}}
\hline\hline
\multicolumn{1}{c}{}&
\multicolumn{2}{c}{Lensing Coefficients}&
\multicolumn{1}{c}{}\\
{$q$} & {$\bar{a}$}&{$\bar{b}$} & {$u_{ps}/R_s$}\\ \hline

\hline                 
                    0  & 1.0000  & -0.40023  & 2.59808\\
\hline                  
                  -0.2 & 1.11557 & -0.289002 & 2.83714\\           
\hline                 
                  -0.4 & 1.27196 & -0.217103 & 3.15179\\
\hline                  
                  -0.6 & 1.48241 & -0.259327 & 3.58493\\
\hline                  
                  -0.8 & 1.71709 & -0.484924 & 4.20506\\ 
\hline
                  -1.0 & 1.82457 & -0.647521 & 5.0838 \\
                  
\hline\hline
	\end{tabular}
	
\caption{Estimates for the strong lensing coefficients $\bar{a}$, $\bar{b}$ and the critical impact parameter $u_{ps}/R_{s}$ for the hairy black hole in Horndeski gravity. The values at  $q=0$ corresponds to the Schwarzschild black hole.  
\label{table1}}
\end{table}   
The strong deflection limit is adopted for the (approximate) analytic calculations, in which the deflection angle is given by Eq.~(\ref{def}), with the coefficients $\bar{a}$ and $\bar{b}$ depending on the specific form of the metric.
The lens equation geometrically governs the connection between the lens observer and the light source. We assume that the source and observer are far from the black hole (lens) and they are perfectly aligned; the equation for small lensing angle reads \cite{Bozza:2001xd}
\begin{equation}\label{lenseq}
\beta=\theta-\frac{D_{LS}}{D_{OS}}\Delta\alpha_{n},
\end{equation}
where  $\Delta\alpha_{n}=\alpha-2n\pi$ is the offset of deflection angle looping over $2n\pi$ and $n$ is an integer. Here, the angular separations between the source and the black hole and observer and source are $\beta$ and $\theta$. $D_{OL}$ and $D_{OS}$ are, respectively, the distance between the observer and the lens and the distance between the observer and the source. Using the Eq.~(\ref{def}) and Eq.~(\ref{lenseq}), the position of the $n$-th relativistic image can be approximated as \cite{Bozza:2002zj} 
\begin{equation}
\theta_n=\theta^0_n+\frac{u_{ps}e_n(\beta-\theta^0_n)D_{OS}}{\bar{a}D_{LS}D_{OL}},
\end{equation}
where
\begin{equation}
e_n=\text{exp}\left({\frac{\bar{b}-2n\pi}{\bar{a}}}\right),
\end{equation}
$\theta^0_n$ are the image positions corresponding to
$\alpha=2n\pi$. As gravitational lensing conserves surface brightness, the magnification is the quotient of the solid angles subtended by the $n$-th image, and the source \cite{Bozza:2002zj,Virbhadra:1999nm,Virbhadra:2008ws}. The magnification of $n$-th relativistic image is thus given by \cite{Bozza:2002zj}
\begin{equation}\label{mag}
\mu_n=\left(\frac{\beta}{\theta} \;  \;\frac{d\beta}{d\theta} \right)^{-1}\Bigg|_{\theta_n ^0} = \frac{u^2_{ps}e_n(1+e_n)D_{OS}}{\bar{a}\beta D_{LS}D^2_{OL}}.
\end{equation}
The first relativistic image is the brightest one, and the magnifications decrease exponentially with $n$. The magnifications are proportional to 1/$D_{OL}^2$, which is a very small factor and thus the relativistic images are very faint, unless $\beta$ has values close to zero, i.e. nearly perfect alignment.

\begin{table*}
	\caption{Estimates for the lensing observables of primary images for black holes in Horndeski gravity and compared with Schwarzschild black $(q=0)$ in GR considering the supermassive black holes Sgr A*, M87*, NGC 4649, and NGC 1332 as lens. The observable $r_{mag}$ does not depend upon the mass or distance of the black hole from the observer.
\label{table2}  
}	
\resizebox{\textwidth}{!}{
 \begin{centering}	
	\begin{tabular}{p{0.8cm} p{1.5cm} p{2cm} p{1.5cm} p{2cm} p{1.5cm} p{2cm} p{1.5cm} p{1.5cm} p{1cm}}
\hline\hline
\multicolumn{1}{c}{}&
\multicolumn{2}{c}{Sgr A*}&
\multicolumn{2}{c}{M87*}& 
\multicolumn{2}{c}{NGC 4649}&
\multicolumn{2}{c}{NGC 1332}\\
{$q$ } & {$\theta_\infty$($\mu$as)} & {$s$($\mu$as)} & {$\theta_\infty $($\mu$as)}  & {$s$($\mu$as)} & {$\theta_\infty $($\mu$as)}  & {$s$($\mu$as)} & {$\theta_\infty $($\mu$as)}  & {$s$($\mu$as)} & $r_{mag}$ \\ \hline\hline

0.0 &  26.3299  & 0.0329517 & 19.782  & 0.0247571 & 14.6615 & 0.0183488 & 7.76719 & 0.00972061 & 6.82188 \\
\hline
-0.2 &  28.7526  & 0.0794533 & 21.6023  & 0.0596944 & 16.0106 & 0.0442427 & 8.4819 & 0.0234384 & 6.11514 \\
\hline
-0.4 &  31.9414  & 0.192717 & 23.998  & 0.144791 & 17.7862 & 0.107312 & 9.42256 & 0.0568506 & 5.36327 \\
\hline
-0.6 &  36.331  & 0.440102 & 27.296  & 0.330656 & 20.2305 & 0.245066 & 10.7175 & 0.129828 & 4.60187 \\
\hline 
-0.8 & 42.6156  & 0.827456 &  32.0177  & 0.62168 & 23.7301 & 0.46076 & 12.5714 & 0.244096 & 3.97294  \\
\hline
-1.0 & 51.5211  & 1.15426  & 38.7086  & 0.867213 & 28.689 & 0.642738 & 15.1985 & 0.340502 & 3.73889 \\

\hline\hline
	\end{tabular}
\end{centering}
}	
\end{table*}

If $\theta_{\infty}$ represents the asymptotic position of a set of images in the limit $n\rightarrow \infty$, we consider  that only  the outermost image $\theta_1$ is resolved as a single image and all the remaining ones are packed together at $\theta_{\infty}$. Having obtained the deflection angle (\ref{def}) and lens equation (\ref{lenseq}) we calculate three observables of relativistic images (cf. Table~\ref{table2}), angular position of the asymptotic relativistic images ($\theta_{\infty}$), angular separation between the outermost and asymptotic relativistic images ($s$) and relative magnification of the outermost relativistic image with other relativistic images ($r_{\text{mag}}$) \cite{Bozza:2002zj,Islam:2021ful}
\begin{align}
\theta_{\infty}& = \frac{u_{ps}} {D_{OL}},\\
s& = \theta_1-\theta_{\infty}=\theta_\infty ~\text{exp}\left({\frac{\bar{b}}{\bar{a}}-\frac{2\pi}{\bar{a}}}\right),\\
r_{\text{mag}}& = \frac{5\pi}{\bar{a}~\text{log}(10)}\label{mag1}.
\end{align}
The strong deflection limit coefficients $\bar{a}$, $\bar{b}$ and
the critical impact parameter $u_{ps}$ can be obtained after 
measuring $s$, $r_{\text{mag}}$ and $\theta_{\infty}$. Then, comparing their values with those predicted by the theoretical models, we can identify the nature of the hairy black holes (lens).

\begin{figure*}[t] 
	\begin{centering}
		\begin{tabular}{cc}
		    \includegraphics[scale=0.75]{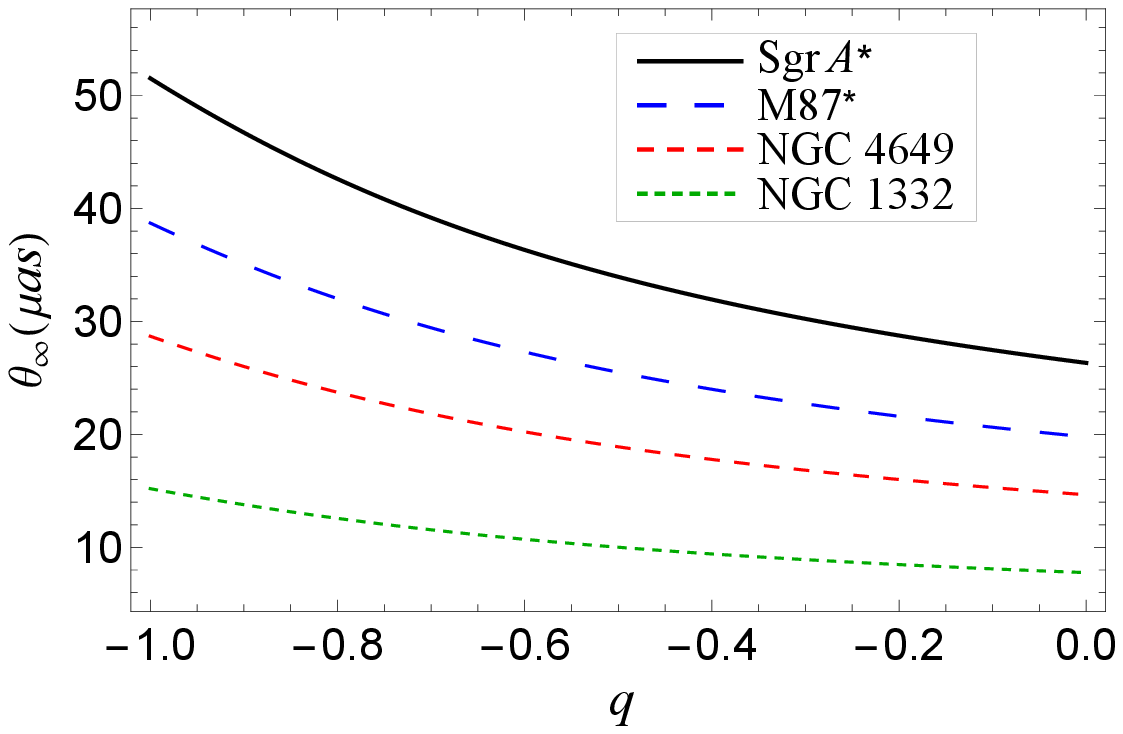}
		    \includegraphics[scale=0.75]{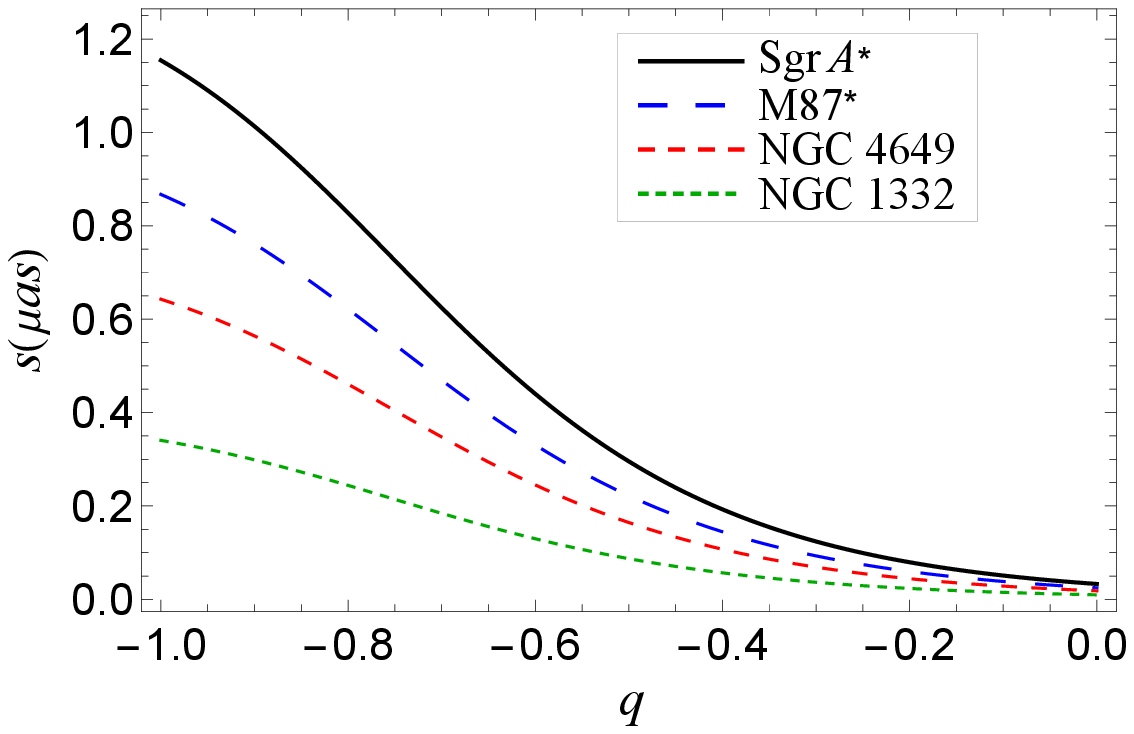}\\
			\end{tabular}
	\end{centering}
	\caption{ The behavior of lensing observables $\theta_{\infty}$ (left), $s$ (right) as a function of hair parameter $q$ in strong field limit by considering the  supermassive black holes at the centres of nearby galaxies as hairy  black holes in Horndeski gravity.}\label{fig7}
\end{figure*}

\begin{figure} 
		      \includegraphics[scale=0.80]{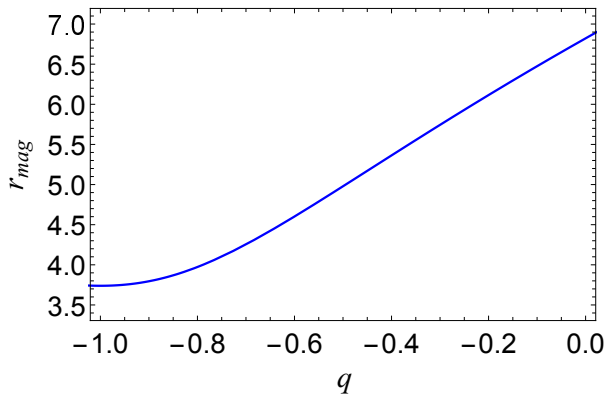}
	\caption{The behavior of strong lensing observable  $r_{mag}$ as function of the parameter $q$. It is independent of the black holes mass or its distance from the observer.}\label{fig8}
\end{figure}

\begingroup
\begin{table*}[tbh!]
	\caption{ Estimation of  time delay for supermassive black holes at the center of nearby  galaxies in the case  Schwarzschild and hairy black holes in Horndeski gravity $(q=-0.5)$. Mass ($M$) and distance ($D_{OL}$) are given in the units of solar mass and Mpc, respectively. Time Delays are expressed in minutes.
	}\label{table3}

	\begin{ruledtabular}
		\begin{tabular}{c c c c c c}  
				Galaxy   &           $M( M_{\odot})$      &          $D_{OL}$ (Mpc)   &     $M/D_{OL}$ & $\Delta T^s_{2,1}(\text{Schw.})$&$\Delta T^s_{2,1}(\text{Hairy})$           \\
			\hline
		
			Milky Way& $  4.3\times 10^6	 $ & $0.0083 $ &       $2.471\times 10^{-11}$ & $11.4968 $ &  $14.8236 $     \\
	
             M87&$ 6.15\times 10^{9} $&$ 16.68 $
&$1.758\times 10^{-11}$& $16443.1 $ &  $21201.2$\\			
		
			 NGC 4472 &$ 2.54\times 10^{9} $&$ 16.72 $
&$7.246\times 10^{-12}$& $6791.11$ &  $8756.28$\\
			
			 NGC 1332 &$ 1.47\times 10^{9} $&$22.66  $
&$3.094\times 10^{-12}$& $3930.29$ &  $5067.61$\\
		
			 NGC 4374 &$ 9.25\times 10^{8} $&$ 18.51 $
&$2.383\times 10^{-12}$& $2473.14$ &  $3188.8$\\
			
			NGC 1399&$ 8.81\times 10^{8} $&$ 20.85 $
&$2.015\times 10^{-12}$& $2355.5$ &  $3037.12$\\
			 
			  NGC 3379 &$ 4.16\times 10^{8} $&$10.70$
&$1.854\times 10^{-12}$& $1112.25$ &  $1434.1$\\
			
			 NGC 4486B &$ 6\times 10^{8} $&$ 16.26 $
&$1.760\times 10^{-12}$ & $ 1604.2$ &  $ 2068.41 $\\
		
			 NGC 1374 &$ 5.90\times 10^{8} $&$ 19.57 $ &$1.438\times 10^{-12}$& $1577.46$ &  $2033.94$\\
			    
			NGC 4649&$ 4.72\times 10^{9} $&$ 16.46 $
&$1.367\times 10^{-12}$& $ 12619.7$ &  $ 16271.5 $\\
		
			NGC 3608 &$  4.65\times 10^{8}  $&$ 22.75  $ &$9.750\times 10^{-13}$& $1243.26$ &  $1603.02$\\
		
			 NGC 3377 &$ 1.78\times 10^{8} $&$ 10.99$
&$7.726\times 10^{-13}$ & $475.913$ &  $613.629$\\
		
			NGC 4697 &$  2.02\times 10^{8}  $&$ 12.54  $ &$7.684\times 10^{-13}$& $540.081$ &  $696.365$\\
			 
			 NGC 5128 &$  5.69\times 10^{7}  $& $3.62   $ &$7.498\times 10^{-13}$& $152.132$ &  $196.154$\\
			
			NGC 1316&$  1.69\times 10^{8}  $&$20.95   $ &$3.848\times 10^{-13}$& $451.85 $ &  $582.603$\\
			
			 NGC 3607 &$ 1.37\times 10^{8} $&$ 22.65  $ &$2.885\times 10^{-13}$& $366.292 $ &  $472.287$\\
			
			NGC 4473 &$  0.90\times 10^{8}  $&$ 15.25  $ &$2.815\times 10^{-13}$& $240.63$ &  $310.262$\\
			
			 NGC 4459 &$ 6.96\times 10^{7} $&$ 16.01  $ &$2.073\times 10^{-13}$ & $186.087 $ &  $239.936$\\
		
			M32 &$ 2.45\times 10^6$ &$ 0.8057 $
&$1.450\times 10^{-13}$ & $6.55048 $ &  $8.44601 $    \\
			
			 NGC 4486A &$ 1.44\times 10^{7} $&$ 18.36  $ &$3.741\times 10^{-14}$ & $38.5008$ &  $49.6419$\\
			 
			NGC 4382 &$  1.30\times 10^{7}  $&$ 17.88 $  &$3.468\times 10^{-14}$& $34.7577 $ &  $44.8156$\\
		
			CYGNUS A &$  2.66\times 10^{9}  $&$ 242.7 $  &$1.4174\times 10^{-15}$& $7111.95 $ &  $9169.96 $\\
		\end{tabular}
	\end{ruledtabular}
\end{table*}
\endgroup

\section{Time Delay in Strong field limit}\label{sec4}
The time difference is caused by the photon taking different paths while winding the black hole, so there is a time delay between different images, which generally depends upon which side of the lens, the images are formed. If we can distinguish the time signals of the first image and other packed images, we can calculate the time delay of two signals \cite{Bozza:2003cp}.   The time spent by the photon winding the black is given by \cite{Bozza:2003cp}
\begin{equation}\label{TD1}
\tilde{T}(u) = \tilde{a} \log\left(\frac{u}{u_{ps}} -1\right) + \tilde{b} +\mathcal{O}(u-u_{ps}).  
\end{equation}

The images are highly demagnified, and the separation between the images is of the order of $\mu$as, so we must at least distinguish the outermost relativistic image from the rest, and we assume the source to be variable, which generally are abundant in all galaxies, otherwise, there is no time delay to measure. The time delay when two images are on the same side is \cite{Bozza:2003cp}
\begin{align}\label{TDS}
\Delta T^s_{n,m} = &2\pi(n-m)\frac{\tilde{a}}{\bar{a}} + 2 \sqrt{\frac{ B(x_{ps})u_{ps}}{A(x_{ps})c}}\nonumber\\
&\left[\text{exp}\left({\frac{\bar{b}-2 m \pi \pm \beta}{2\bar{a}}}\right)-\text{exp}\left({\frac{\bar{b}-2 n \pi \pm \beta}{2\bar{a}}}\right)\right].
\end{align} 
The upper sign before $\beta$ signifies that both the images are on the same side of the source and the lower sign if the images are on the other side. When the images are on the opposite sides of lens, the time dilation between $m^{\text{th}}$ and $n^{\text{th}}$ relativistic image is give
\begin{align}\label{TDO}
\Delta T^o_{n,m} = &[2\pi(n-m)-2\beta]\frac{\tilde{a}}{\bar{a}} + 2 \sqrt{\frac{ B(x_{ps})u_{ps}}{A(x_{ps})c}}\nonumber\\
&\left[ \text{exp}\left({\frac{\bar{b}-2 m \pi + \beta}{2\bar{a}}}\right)-\text{exp}\left({\frac{\bar{b}-2 n \pi + \beta}{2\bar{a}}}\right)\right].
\end{align} 
The contribution of second term in Eq.~(\ref{TDS}) and Eq.~(\ref{TDO}) is very small. For spherically symmetric black holes, the time delay between the first and second relativistic image is given by 
\begin{equation}\label{deltaT}
\Delta T^s_{2,1} = 2\pi u_{ps} = 2\pi D_{OL} \theta_{\infty}.
\end{equation}  
Using Eq.~(\ref{deltaT}), if we can measure the time delay with an accuracy of $ 5\% $ and critical impact parameter with negligible error, we can get the distance of the black hole with an accuracy of $5\%$. In Table~\ref{table3}, we compare the values of time delay between the first and second relativistic image considering the black hole at the center of several nearby galaxies to be Schwarzschild black hole and hairy black hole in Horndeski gravity at $q = -0.5$.

\begingroup
\begin{table*}
	\caption{
		{\bf Image positions  of first and second order primary and secondary images due to lensing by Sgr A* with $d=D_{LS}/D_{OS}=0.5$}: GR and Horndeski Gravity $(q=-0.5)$ predictions for angular positions $\theta$ of primary ($p$)   and secondary images ($s$)  are given for different values of angular source position $\beta$. {\bf (a)} All angles are in $\mu${\em as}. {\bf (b)} We have used $M_{\text{Sgr A*}}= 4.3\times 10^6	\, {\rm m}$, $D_{OL}= 8.3 \times 10^{6} \, {\rm pc}$.
	}\label{table4}
	\begin{ruledtabular}
		\begin{tabular}{l cccc cccc}
			\multicolumn{1}{c}{$\beta$}&
			\multicolumn{4}{c}{ Horndeski Gravity}&
			\multicolumn{4}{c}{General relativity}\\
&$\theta_{1p,{\rm HG}} $&$\theta_{2p,{\rm HG}}$&$\theta_{1s,{\rm HG}}$&$\theta_{2s,{\rm HG}}$&$\theta_{1p,{\rm GR}} $&$\theta_{2p,{\rm GR}}$&$\theta_{1s,{\rm GR}}$&$\theta_{2s,{\rm GR}}$\\
			\hline
		
            $0$& 34.2444 & 33.952 & -34.2444 & -33.952 & 26.3628 & 26.3299 & -26.3628 & -26.3299 \\
			$10^{0} $& 34.2444 & 33.952 & -34.2444 & -33.952 & 26.3628 & 26.3299 & -26.3628 & -26.3299 \\
			$10^{1} $ & 34.2445 & 33.952 & -34.2444 & -33.952 & 26.3628 & 26.3299 & -26.3628 & -26.3299 \\
			$10^{2} $ & 34.2446 & 33.952 & -34.2442 & -33.952 & 26.3628 & 26.3299 & -26.3628 & -26.3299 \\
			$10^{3} $ & 34.2465 & 33.952 & -34.2423 & -33.952 & 26.3631 & 26.3299 & -26.3625 & -26.3299 \\
			$10^{4} $  & 34.2653 & 33.9522 & -34.2235 & -33.9518 & 26.366 & 26.3299 & -26.3596 & -26.3299 \\
		\end{tabular}
	\end{ruledtabular}
\end{table*}
\endgroup

\begingroup
\begin{table*}
	\caption{
		{\bf Image positions  of first and second order primary and secondary images due to lensing by M87* with $d=D_{LS}/D_{OS}=0.5$}: GR and Horndeski Gravity $(q=-0.5)$ predictions for angular positions $\theta$ of primary ($p$)   and secondary images ($s$)  are given for different values of angular source position $\beta$. {\bf (a)} All angles are in $\mu${\em as}. {\bf (b)} We have used $M_{\text{M87*}}=  6.5 \times 10^9 \, {\rm m}$, $D_{OL}=  16.8  \times 10^6 \, {\rm pc}$.
	}\label{table5}
	\begin{ruledtabular}
		\begin{tabular}{l cccc cccc}
			\multicolumn{1}{c}{$\beta$}&
			\multicolumn{4}{c}{ Horndeski Gravity}&
			\multicolumn{4}{c}{General relativity}\\
&$\theta_{1p,{\rm HG}} $&$\theta_{2p,{\rm HG}}$&$\theta_{1s,{\rm HG}}$&$\theta_{2s,{\rm HG}}$&$\theta_{1p,{\rm GR}} $&$\theta_{2p,{\rm GR}}$&$\theta_{1s,{\rm GR}}$&$\theta_{2s,{\rm GR}}$\\
			\hline
	
            $0$& 25.7284 & 25.5087 & -25.7284 & -25.5087 & 19.8068 & 19.7821 & -19.8068 & -19.7821 \\
			$10^{0} $&  25.7284 & 25.5087 & -25.7284 & -25.5087 & 19.8068 & 19.7821 & -19.8068 & -19.7821 \\
			$10^{1} $ &  25.7284 & 25.5087 & -25.7283 & -25.5087 & 19.8068 & 19.7821 & -19.8068 & -19.7821 \\
			$10^{2} $ & 25.7285 & 25.5087 & -25.7282 & -25.5087 & 19.8068 & 19.7821 & -19.8068 & -19.7821 \\
			$10^{3} $ &  25.7299 & 25.5087 & -25.7268 & -25.5086 & 19.807 & 19.7821 & -19.8065 & -19.7821 \\
			$10^{4} $  &  25.7441 & 25.5088 & -25.7127 & -25.5085 & 19.8092 & 19.7821 & -19.8044 & -19.7821 \\
		\end{tabular}
	\end{ruledtabular}
\end{table*}
\endgroup

\begin{table*}[t]
	\caption{
		{\bf Magnifications  of first order and second order relativistic images due to lensing by Sgr A* with $d=D_{LS}/D_{OS}=0.5$}: GR and Horndeski Gravity $(q=-0.5)$ predictions for magnifications $\mu_n$  is given for different values of angular source position $\beta$. {\bf (a)} $1p$ and $1s$ refer to first order relativistic images on the same side as primary and secondary images, respectively. {\bf (b)}  We have used $M_{\text{Sgr A*}}= 4.3\times 10^6	\, {\rm m}$, $D_{OL}= 8.3 \times 10^{6} \, {\rm pc}$ {\bf (c)} Angular positions of first order relativistic images in GR and Horndeski Gravity are, respectively, $\theta_{1p,{\rm GR}}\approx -\theta_{1s,{\rm GR}}\approx 26.3628  \mu as$ and $\theta_{1p,{\rm HG}}\approx -\theta_{1s,{\rm HG}}\approx 34.244  \mu as$ and are highly insensitive to the angular source position $\beta$.
	}\label{table6}
\resizebox{\textwidth}{!}{
 \begin{centering}	
	\begin{tabular}{p{0.8cm} p{2.5cm} p{2.5cm} p{2.7cm} p{2.7cm} p{2.5cm} p{2.2cm} p{2.7cm} p{2.2cm}}
\hline\hline
	\multicolumn{1}{c}{$\beta$}&
			\multicolumn{4}{c}{Horndeski Gravity}&
			\multicolumn{4}{c}{General relativity}\\
    &  $\mu_{1p,{\rm HG}} $ & $ \mu_{2p,{\rm HG}}$&$\mu_{1s,{\rm HG}}$&$\mu_{2s,{\rm HG}}$&$\mu_{1p,{\rm GR}}$&$\mu_{2p,{\rm GR}}$&$\mu_{1s,{\rm GR}}$&$\mu_{2s,{\rm GR}}$\\
	\hline 	\hline

		
	    	$10^{0}  $&$7.2458 \times 10^{-11}$&$ 7.32635  \times	10^{-13} $&$-7.2458 \times 10^{-11} $&$ -7.32635  \times	10^{-13} $&$8.52495 \times 10^{-12}  $&$1.59  \times	10^{-14} $&$-8.52495 \times 10^{-12}  $&$-1.59  \times	10^{-14}  $\\
			
			$10^{1}  $&$7.2458 \times 10^{-12}$&$ 7.32635  \times	10^{-14} $&$-7.2458 \times 10^{-12} $&$ -7.32635  \times	10^{-14} $&$8.52495 \times 10^{-13}$&$1.59  \times	10^{-15} $&$-8.52495 \times 10^{-13}  $&$- 1.59  \times	10^{-15}	$\\
			
			$10^{2}  $&$ 7.2458 \times 10^{-13} $&$ 7.32635 \times	10^{-15} $&$-7.2458 \times 10^{-13} $&$ -7.32635  \times	10^{-15} $&$8.52495 \times 10^{-14} $&$1.59  \times	10^{-16} $&$-8.52495 \times 10^{-14} $&$ -1.59  \times	10^{-16}	$\\
			
			$10^{3}  $&$7.2458 \times 10^{-14} $&$ 7.32635  \times	10^{-16} $&$-7.2458 \times 10^{-14}$&$ -7.32635  \times	10^{-16} $&$8.52495 \times 10^{-15}  $&$ 1.59  \times	10^{-17} $&$-8.52495 \times 10^{-15}  $&$ -1.59  \times	10^{-17} 		$\\
			
			$10^{4}  $&$7.2458 \times 10^{-15} $&$  7.32635  \times	10^{-17} $&$-7.2458 \times 10^{-15} $&$ -7.32635  \times	10^{-17} $&$8.52495 \times 10^{-16} $&$1.59  \times	10^{-18}$&$-8.52495 \times 10^{-16}  $&$ -1.59  \times	10^{-18} 	$\\
\hline\hline
	\end{tabular}
\end{centering}
}	
\end{table*}

\begin{table*}[t]
	\caption{
		{\bf Magnifications  of first and second order relativistic images due to lensing by M87* with $d=D_{LS}/D_{OS}=0.5$}: GR and Horndeski Gravity $(q=-0.5)$ predictions for magnifications $\mu_n$ is given for different values of angular source position $\beta$. {\bf (a)} $1p$ and $1s$ refer to first order relativistic images on the same side as primary and secondary images, respectively. {\bf (b)}  We have used $M_{\text{M87*}}=  6.5 \times 10^9 \, {\rm m}$, $D_{OL}=  16.8  \times 10^6 \, {\rm pc}$. {\bf (c)} Angular positions of first order relativistic images in GR and Horndeski Gravity are, respectively, $\theta_{1p,{\rm GR}}\approx -\theta_{1s,{\rm GR}}\approx 19.8068  \mu as$ and $\theta_{1p,{\rm HG}}\approx -\theta_{1s,{\rm HG}}\approx 25.7284 \mu as$ and are highly insensitive to the angular source position $\beta$.
	}\label{table7}
\resizebox{\textwidth}{!}{
 \begin{centering}	
	\begin{tabular}{p{0.8cm} p{2.5cm} p{2.5cm} p{2.7cm} p{2.7cm} p{2.5cm} p{2.5cm} p{2.7cm} p{2.5cm}}
\hline\hline
	\multicolumn{1}{c}{$\beta$}&
			\multicolumn{4}{c}{Horndeski Gravity}&
			\multicolumn{4}{c}{General relativity}\\
    &  $\mu_{1p,{\rm HG}} $ & $ \mu_{2p,{\rm HG}}$&$\mu_{1s,{\rm HG}}$&$\mu_{2s,{\rm HG}}$&$\mu_{1p,{\rm GR}}$&$\mu_{2p,{\rm GR}}$&$\mu_{1s,{\rm GR}}$&$\mu_{2s,{\rm GR}}$\\
	\hline 	\hline

		
			$10^{0}$  & $4.04123 \times 10^{-11}$ & $4.08616  \times	10^{-13}$ &$-4.04123  \times 10^{-11} $&$ -4.08616  \times	10^{-13} $&$4.75466  \times 10^{-12}  $&$8.86797 \times	10^{-15} $&$-4.75466 \times	10^{-12}  $&$-8.86797  \times	10^{-15}  $\\
			
			$10^{1}  $&$4.04123\times10^{-12}$&$ 4.08616  \times	10^{-14} $&$ -4.04123\times10^{-12}$&$ -4.08616  \times	10^{-14} $&$4.75466\times10^{-13}$&$8.86797 \times	10^{-16} $&$-4.75466\times10^{-13}  $&$ -8.86797  \times	10^{-16}	$\\
			
			$10^{2}  $&$4.04123\times10^{-13} $&$ 4.08616  \times	10^{-15} $&$ -4.04123\times10^{-13}$&$ -4.08616  \times	10^{-15} $&$ 4.75466\times10^{-14} $&$8.86797  \times	10^{-17} $&$-4.75466\times10^{-14} $&$ -8.86797  \times	10^{-17}	$\\
			
			$10^{3}  $&$4.04123\times10^{-14} $&$ 4.08616  \times	10^{-16} $&$-4.04123\times10^{-14}$&$-4.08616  \times	10^{-16} $&$ 4.75466\times10^{-15}  $&$ 8.86797 \times	10^{-18} $&$-4.75466 \times 10^{-15}  $&$ -8.86797  \times	10^{-18}		$\\
			
			$10^{4}  $&$4.04123\times10^{-15} $&$ 4.08616 \times	10^{-17} $&$-4.04123\times10^{-15} $&$ -4.08616  \times	10^{-17} $&$4.75466\times10^{-16} $&$8.86797 \times	10^{-19} $&$-4.75466\times10^{-16}  $&$ -8.86797 \times	10^{-19}  		$\\
\hline\hline
	\end{tabular}
\end{centering}
}	
\end{table*}
\section{Gravitational lensing parameters for supermassive black holes}\label{sec5}
Next, we model the supermassive black holes in the nearby galaxies especially Sgr A*, M87*, NGC4649 and NGC1332 as the hairy black hole to estimate and compare the observables with those of Schwarzschild black hole of GR. The mass and distance from earth for Sgr A* \cite{Do:2019vob}  are $M \approx 4.3 \times 10^6 M_{\odot}$, $D_{OL}  \approx  8.35\text{kpc}$, for  M87* \cite{Akiyama:2019cqa} are  $M  \approx  6.5 \times 10^9 M_{\odot}$, $D_{OL}  \approx  16.8 \text{Mpc}$, for  NGC4649 \cite{Kormendy:2013} are  $M  \approx  4.72 \times 10^9 M_{\odot}$, $D_{OL}  \approx  16.46 \text{Mpc}$ and for  NGC1332 \cite{Kormendy:2013} are  $M  \approx  2.54 \times 10^9 M_{\odot}$, $D_{OL}  \approx  16.72 \text{Mpc}$. Using Eq.~(\ref{lenseq}), we compute the angular positions for first and second order relativistic primary and secondary images; the images on the same and opposite sides of the source respectively, taking $d=D_{LS}/D_{OS}=0.5$, for Sgr A* and M87* black holes. First (Second) order relativistic images are produced after the light winds, once (twice) around the black hole before reaching the observer. From the results in Table \ref{table4} and \ref{table5} it shows that in  the Horndeski gravity the angular positions of images  are larger than their corresponding values in GR and are very insensitive to the position of the source $\beta$. The angular positions  $\theta_{1p} > |\theta_{1s}|$ for higher values of $\beta$, however, for small values of $\beta$  the values are  extremely close. The same is true for any pair of second or higher order relativistic images. In Horndeski gravity the first and second order primary images are   about 7.8 and 7.6 $\mu$as larger than their corresponding values in GR  at $q=-0.5$, an effect too tiny to be observed with today's telescopes, especially since these relativistic images are highly demagnified. However the deviation becomes significant at higher magnitude of $q$ and the next generation telescope (ngEHT) which  renders these observables, the deviations could be used to test Horndeski gravity. The characteristic observables which include position of the innermost image $\theta_{\infty}$ and the separation $s$, are depicted in Fig.~\ref{fig7} and tabulated in Table \ref{table2}. Considering the Sgr A* and M87* as the lens we find that these observables vary rapidly with parameter $q$; with $\theta_{\infty}$ ranging in between 26.33 - 51.52 $\mu$as  for Sgr A*  and 19.78 - 38.71 $\mu$as for M87* and the latter are consistent with the EHT measured diameters of M87* shadow $42\pm 3~\mu\text{as}$. The deviation from their GR counterpart are quite significant and can reach as much as  25.1912 $\mu$as for Sgr A* and 18.92$\mu$as for M87*. Further, the seperation $s$ due to hairy black holes for Sgr A* and M87* range between 0.0329-1.15426 $\mu$as and 0.024-0.867 $\mu$as, respectively. We also obtain these results for NGC 4649 and NGC 1332 and found that the deviation is also of the order of $\mathcal{O}(\mu)$as. As the deviation of position of the images are quite significant at higher magnitude of $q$ and  these images could  be resolved, it is possible to measure the brightness difference. The relative  magnification  of the first and second order images of Table \ref{table4} and \ref{table5} are tabulated in Table \ref{table6} and \ref{table7}  using Eq.~(\ref{mag}) for black holes in  GR and Horndeski gravity at $(q=-0.5)$. The first order images in Horndeski gravity are highly magnified than the second order images as well the corresponding images in GR. However, the ratio of the flux of the first image to the all other images rapidly decreases with decreasing  $q$ implying that the Schwarzschild images are bighter than the hairy black holes in Horndeski gravity (cf. Fig.~\ref{fig7}). Finally, we  collect some updated data of 14 supermassive black holes whose masses and distances considerably  differ  from that of Sgr A*, to calculate the  time delays between the first and second order relativistic primary images $\Delta T^s_{2,1}$ in Table \ref{table3}. For Sgr A* and M87*, the time delay can reach $\sim14.82$~min and $\sim21201.2$~min at $q=-0.5$ and hence deviate from their corresponding black hole in GR by $\sim3.32$~min and $\sim4758.1$~min. Although these deviations are insignificant for Sgr A*  but for M87* and other black holes these  are sufficient values  to test and compare  the Horndeski gravity from GR.     
\section{Conclusion}\label{sec6}
 One of the most renowned scalar-tensor theories is Horndeski gravity, in which scalar fields constitute additional degrees of freedom and is the most general four-dimensional scalar-tensor theory with equations of motion containing second-order derivatives of the dynamical fields.   We have investigated gravitational lensing effects of spherically symmetric hairy black holes in Horndeski gravity having additional paramter $q$ and also  consider the predictions of Horndeski gravity for lensing effects by supermassive black
holes Sgr A*, M87* and 21 others in comparison with GR. 
We examined the effect of the parameter $q$ on the light deflection angle $\alpha_{D}$, strong lensing coefficients $\bar{a}$, $\bar{b}$ and lensing observables $\theta_{\infty}$, $s$, $r_{\text{mag}}$, $u_{ps}$ and time delay $\Delta T^s_{2,1}$ in the strong-field regime, due to the hairy black hole in Horndeski gravity and compared them to the Schwarzschild ($q=0$) black hole of GR. We  found that $\bar{a}$ and $u_{ps}$ increase monotonically with increasing magnitude of $q$ while $\bar{b}$ first increases, reaching its maximum at $q \approx -0.45$, and then decreases. The numerical values, in our selected range of $q$, are positive for $\bar{a}$ while negative for $\bar{b}$. We found that deflection angle $\alpha_D$, for fixed impact parameter $u$, is greater for the hairy black hole in Horndeski gravity when compared to the Schwarzschild black hole and  increases with the increasing magnitude of  $q$. Also, the photon sphere radius $x_{ps}$ increases with decreasing  $q$, making bigger photon spheres in the hairy black hole in Horndeski gravity when compared to the Schwarzschild black holes of GR.

We  calculated lensing observables $\theta_{\infty} $, $s$ and $r_{\text{mag}}$ of the relativistic images for supermassive black holes, namely, Sgr A*, M87, NGC 4649 and NGC 1332 by considering the spacetime to be described by the hairy black hole in Horndeski gravity. In its predictions for gravitational lensing due to supermassive black holes, Horndeski gravity exhibits potentially observable departures from GR. The presence of paramter $q$ rapidly increases $\theta_{\infty} $ and $s$  when compared to the Schwarzschild ($q=0$) black hole. We observe that  $\theta_{\infty}$ ranges between 26.33 - 51.52 $\mu$as for Sgr A* and its deviation from its GR counterpart can reach as much as  25.1912 $\mu$as while as for M87* it ranges between  19.78 - 38.71 $\mu$as and deviation is as high as  18.92$\mu$as. On the other hand the seperation $s$ due to hairy black holes for Sgr A* and M87* range between 0.0329-1.15426 $\mu$as and 0.024-0.867 $\mu$as, respectively. In the limit, $q \rightarrow 0$, our results reduce exactly to the Schwarzschild black hole results. The angular positions of images though are very insensitive to the position of the source $\beta$ with $\theta_{1p} > |\theta_{1s}|$ at  higher values. $\theta_{1p}$  and $\theta_{2p}$, which are the  angular positions of  first and second order primary images are 7.8 $\mu$as and 7.6 $\mu$as larger than their corresponding values in GR  at $q=-0.5$.  The first order images in Horndeski gravity are highly magnified than  the corresponding images in GR.  However, $r_{\text{mag}}$ rapidly decreases with $q$ suggesting  that the Schwarzschild images are brighter than the hairy black holes in Horndeski gravity. 
Finally, the time delay of the first and second order images for hairy black holes in Horndeski gravity 
is significantly larger (e.g.  $\sim4758.1$~min for M87* ) than the GR counterparts  for astronomical measurements, provided we have enough angular resolution separating two relativistic images, except for Sgr A* for which the deviation is $\sim3.32$~min; an effect too tiny to be detected  by EHT. 

Many interesting avenues are amenable for future work from the  hairy black holes in Horndeski gravity; most importantly is to consider the rotating black holes as there is good observational evidence
that many black holes are spinning.  Our results will certainly be different in this case, and likely substantively  for near-extremal solutions. Also, it will be intresting to analyze the relationship between the null geodesics and thermodynamic phase transition in AdS background in the context photon sphere.  Work on these problems is in progress. Further, gravitational lensing in the strong field may open fascinating perspectives for testing modified theories of gravity and estimating the parameters associated with the supermassive black holes.

\section{Acknowledgments} 
 J.K. would like to thank CSIR for providing JRF.

\end{document}